\documentclass[twocolumn,showpacs,aps,prl,superscriptaddress]{revtex4}

\usepackage{graphicx}
\usepackage{dcolumn}
\usepackage{amsmath}
\usepackage{epsfig}
\usepackage{psfrag}

\newcommand{\Ks}{K_S^0}
\newcommand{\Kt}{K^{*0}}
\newcommand{\pipm}{\pi^+ \pi^-}

\newcommand{\GeVc}{{\rm GeV}/c}
\newcommand{\GeVcc}{{\rm GeV}/c^2}

\newcommand{\MeVcc}{{\rm MeV}/c^2}

\newcommand{\um}{\mu {\rm m}}
\newcommand{\BR}{{\rm BR}}

\begin{document}


\title{A Study of $B^0 \to J/\psi K^{(*)0} \pipm$ Decays with the
Collider Detector at Fermilab}

\newcommand{\affilAB} {\affiliation{Institute of Physics, Academia Sinica, Taipei, Taiwan 11529, 
	Republic of China}}
\newcommand{\affilAC} {\affiliation{Argonne National Laboratory, Argonne, Illinois 60439}}
\newcommand{\affilAD} {\affiliation{Istituto Nazionale di Fisica Nucleare, University of Bologna,
	I-40127 Bologna, Italy}}
\newcommand{\affilAE} {\affiliation{Brandeis University, Waltham, Massachusetts 02254}}
\newcommand{\affilAF} {\affiliation{University of California at Davis, Davis, California  95616}}
\newcommand{\affilAG} {\affiliation{University of California at Los Angeles, Los 
	Angeles, California  90024}}
\newcommand{\affilAH} {\affiliation{Instituto de Fisica de Cantabria, CSIC-University of Cantabria, 
	39005 Santander, Spain}}
\newcommand{\affilAI} {\affiliation{Enrico Fermi Institute, University of Chicago, Chicago, 
	Illinois 60637}}
\newcommand{\affilAJ} {\affiliation{Joint Institute for Nuclear Research, RU-141980 Dubna, Russia}}
\newcommand{\affilBA} {\affiliation{Duke University, Durham, North Carolina  27708}}
\newcommand{\affilBB} {\affiliation{Fermi National Accelerator Laboratory, Batavia, Illinois 60510}}
\newcommand{\affilBC} {\affiliation{University of Florida, Gainesville, Florida  32611}}
\newcommand{\affilBD} {\affiliation{Laboratori Nazionali di Frascati, Istituto Nazionale di Fisica
               Nucleare, I-00044 Frascati, Italy}}
\newcommand{\affilBE} {\affiliation{University of Geneva, CH-1211 Geneva 4, Switzerland}}
\newcommand{\affilBF} {\affiliation{Glasgow University, Glasgow G12 8QQ, United Kingdom}}
\newcommand{\affilBG} {\affiliation{Harvard University, Cambridge, Massachusetts 02138}}
\newcommand{\affilBH} {\affiliation{Hiroshima University, Higashi-Hiroshima 724, Japan}}
\newcommand{\affilBI} {\affiliation{University of Illinois, Urbana, Illinois 61801}}
\newcommand{\affilBJ} {\affiliation{The Johns Hopkins University, Baltimore, Maryland 21218}}
\newcommand{\affilCA} {\affiliation{Institut f\"{u}r Experimentelle Kernphysik, 
	Universit\"{a}t Karlsruhe, 76128 Karlsruhe, Germany}}
\newcommand{\affilCB} {\affiliation{Center for High Energy Physics: Kyungpook National
	University, Taegu 702-701; Seoul National University, Seoul 151-742; and
	SungKyunKwan University, Suwon 440-746; Korea}}
\newcommand{\affilCC} {\affiliation{High Energy Accelerator Research Organization (KEK), Tsukuba, 
	Ibaraki 305, Japan}}
\newcommand{\affilCD} {\affiliation{Ernest Orlando Lawrence Berkeley National Laboratory, 
	Berkeley, California 94720}}
\newcommand{\affilCE} {\affiliation{Massachusetts Institute of Technology, Cambridge,
	Massachusetts  02139}}
\newcommand{\affilCF} {\affiliation{University of Michigan, Ann Arbor, Michigan 48109}}
\newcommand{\affilCG} {\affiliation{Michigan State University, East Lansing, Michigan  48824}}
\newcommand{\affilCH} {\affiliation{University of New Mexico, Albuquerque, New Mexico 87131}}
\newcommand{\affilCI} {\affiliation{The Ohio State University, Columbus, Ohio  43210}}
\newcommand{\affilCJ} {\affiliation{Osaka City University, Osaka 588, Japan}}
\newcommand{\affilDA} {\affiliation{University of Oxford, Oxford OX1 3RH, United Kingdom}}
\newcommand{\affilDB} {\affiliation{Universita di Padova, Istituto Nazionale di Fisica 
        Nucleare, Sezione di Padova, I-35131 Padova, Italy}}
\newcommand{\affilDC} {\affiliation{University of Pennsylvania, Philadelphia, 
        Pennsylvania 19104}}
\newcommand{\affilDD} {\affiliation{Istituto Nazionale di Fisica Nucleare, University and Scuola
	Normale Superiore of Pisa, I-56100 Pisa, Italy}}
\newcommand{\affilDE} {\affiliation{University of Pittsburgh, Pittsburgh, Pennsylvania 15260}}
\newcommand{\affilDF} {\affiliation{Purdue University, West Lafayette, Indiana 47907}}
\newcommand{\affilDG} {\affiliation{University of Rochester, Rochester, New York 14627}}
\newcommand{\affilDH} {\affiliation{Rockefeller University, New York, New York 10021}}
\newcommand{\affilDI} {\affiliation{Rutgers University, Piscataway, New Jersey 08855}}
\newcommand{\affilDJ} {\affiliation{Texas A\&M University, College Station, Texas 77843}}
\newcommand{\affilEA} {\affiliation{Texas Tech University, Lubbock, Texas 79409}}
\newcommand{\affilEB} {\affiliation{Institute of Particle Physics, University of Toronto, Toronto
	M5S 1A7, Canada}}
\newcommand{\affilEC} {\affiliation{Istituto Nazionale di Fisica Nucleare, University of Trieste/
	Udine, Italy}}
\newcommand{\affilED} {\affiliation{University of Tsukuba, Tsukuba, Ibaraki 305, Japan}}
\newcommand{\affilEE} {\affiliation{Tufts University, Medford, Massachusetts 02155}}
\newcommand{\affilEF} {\affiliation{Waseda University, Tokyo 169, Japan}}
\newcommand{\affilEG} {\affiliation{University of Wisconsin, Madison, Wisconsin 53706}}
\newcommand{\affilEH} {\affiliation{Yale University, New Haven, Connecticut 06520}}

\author{T.~Affolder}	\affilCD
\author{H.~Akimoto}	\affilEF
\author{A.~Akopian}	\affilDH
\author{M.~G.~Albrow}	\affilBB
\author{P.~Amaral}	\affilAI  
\author{D.~Amidei}	\affilCF
\author{K.~Anikeev}	\affilCE
\author{J.~Antos}	\affilAB 
\author{G.~Apollinari}	\affilBB
\author{T.~Arisawa}	\affilEF
% ...
\collaboration{CDF Collaboration}
\noaffiliation

\date{27 July 2001}

\begin{abstract}
We report a study of the decays $B^0 \to J/\psi K^{(*)0} \pipm$,
which involve the creation of a $u \bar u$ or
$d \bar d$ quark pair in addition to a $\bar b \to \bar c(c \bar s)$ decay.
The data sample consists of 110 pb$^{-1}$ of $p \bar p$ collisions
at $\sqrt{s} = 1.8$ TeV collected by the CDF detector at the
Fermilab Tevatron collider during 1992-1995.  We measure the branching ratios
to be $\BR(B^0 \to J/\psi K^{*0} \pipm) = (8.0 \pm 2.2 \pm 1.5) \times 10^{-4}$
and $\BR(B^0 \to J/\psi K^0 \pipm) = (1.1 \pm 0.4 \pm 0.2) \times 10^{-3}$.
Contributions to these decays are seen from
$\psi(2S) K^{(*)0}$, $J/\psi K^0 \rho^0$, $J/\psi K^{*+} \pi^-$, and
$J/\psi K_1(1270)$.
\end{abstract}


\maketitle



\begin{figure}[ht]
\begin{center}
\epsfig{file=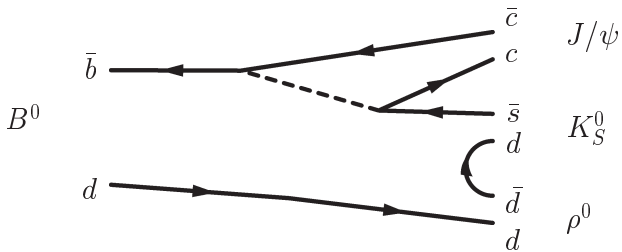, width=3.25in}
\end{center}
\caption{\label{fig:feynman}
Example of a $\bar b \to \bar c(c \bar s)$ decay with
a $d \bar d$ quark pair and spectator.}
\end{figure}

The measured inclusive branching ratio for $B \to J/\psi X$ of
$(1.16 \pm 0.10)\%$ is considerably larger than the sum of the
individual branching ratios of the known exclusively reconstructed
decays \cite{PDG}.
One possible source of the missing decay modes is a class of decays
in which a quark pair is created in addition to a
$\bar b \to \bar c(c \bar s)$ decay.
The CLEO collaboration recently reported observing one such mode,
$B \to J/\psi \phi K$ \cite{CLEO_psiphik},
which involves an $s\bar s$ quark pair.
This analysis studies similar $B^0$ decays that involve
$u \bar u$ or $d \bar d$ quark pairs, an example of which is shown
in Fig.~\ref{fig:feynman}.

These modes are potentially useful for $CP$ violation measurements.
For example, $J/\psi \Ks \rho^0$ is accessible from both $B^0$ and $\bar B^0$
which allows $CP$ violation due to interference between decays with and
without mixing.
An angular analysis of the decay products could separate
the $CP$-even and -odd contributions.
If the factorization hypothesis \cite{factorization} holds,
a time dependent analysis of
the angular correlations could measure $\cos(2\beta)$ in a manner
similar to $B^0 \to J/\psi (K^{*0} \to \Ks \pi^0)$ \cite{angularCP}.


This analysis was performed using $p \bar p$ collisions recorded with
the CDF detector, which is described in detail
elsewhere \cite{CDF}.  For this analysis the important components
are the Silicon Vertex Detector (SVX), the Central Tracking Chamber (CTC),
and the central muon systems.
The SVX provides a track impact parameter
resolution of $\sim (13 + 40/p_T) \; \um$, where $p_T$ (in $\GeVc$) is
the component of the momentum transverse to the $p \bar p$ collision
axis (the $z$ axis) \cite{CDF_coords}.
The CTC is a drift chamber whose charged particle
momentum resolution is $\delta p_T/p_T^2 \sim 0.001 / (\GeVc)$.
Two muon systems separated by 60 cm of steel cover the region
$|\eta| < 0.6$ for muons with $p_T > 1.4 \; \GeVc$.
Each of these central muon systems consists of four layers of planar
drift chambers.  The inner system is separated from the interaction
point by an average of five interaction lengths of material.
An extension to the central muon systems covers $0.6 < |\eta| < 1.0$.

This analysis uses a three-level dimuon trigger.
The first level selects events with two separate sets of
at least three linked hits in the muon chambers that cover $|\eta| < 1.0$.
The second level requires drift
chamber tracks with $p_T > 2.0 \; \GeVc$ which extrapolate to the
linked hits in the muon chambers.  The third level accepts
$J/\psi \to \mu^+ \mu^-$
candidates with a reconstructed invariant mass between 2.8 and 3.4
$\GeVcc$.  In addition to this trigger path, approximately
$10\%$ of the events which pass the level 3 dimuon trigger
come from single muon triggers at levels 1 and 2
with muon $p_T$ thresholds of either 7.5 $\GeVc$ or 12 $\GeVc$.

The offline analysis reconstructs $J/\psi \to \mu^+ \mu^-$
candidates but does not require them to be the same candidate which
passed the trigger.  The offline reconstruction only uses muons which
intersect both of the muon systems that cover $|\eta| < 0.6$.


A ratio of branching ratios is measured
between a signal mode $B^0 \to J/\psi K^{(*)0} \pipm$ and a
well established reference mode $B^0 \to J/\psi K^{(*)0}$ \cite{PDG}:
\begin{equation}
\label{eq:refmode}
\frac{\BR_{sig}}{\BR_{ref}} =
	\frac{\epsilon_{ref}}{\epsilon_{sig}} \frac{N_{sig}}{N_{ref}}
\end{equation}
Many systematic uncertainties cancel in this ratio.
The ratio of efficiencies $R_\epsilon = \epsilon_{ref} / \epsilon_{sig}$
is determined with a Monte Carlo simulation.  The number of signal
and reference events ($N_{sig}$ and $N_{ref}$)
are measured in the data while applying similar selection criteria
to both signal and reference
decay modes.  The only selection criteria which differ are those
placed upon the two extra pions of the signal events for which there
are no equivalents in the reference decays.

From the dimuon trigger events, $B^0$ decay candidates are selected which
satisfy the basic topology of the decays of interest.  The reference
modes are reconstructed by combining a $J/\psi \to \mu^+ \mu^-$ candidate
with either a $\Ks \to \pipm$ or $\Kt \to K^+ \pi^-$ candidate.
$\Ks$ candidates are required to have an invariant mass
between 485 and 510 $\MeVcc$ and to point back to the $J/\psi$ decay
vertex.  Additionally, the $\Ks$ candidates' decay vertices are
required to have a positive displacement in the $xy$ plane
from the $J/\psi$ decay vertex with at least $5 \sigma$ significance.
$\Kt$ candidates are required to have an invariant mass between
820 and 970 $\MeVcc$ and to originate from the $J/\psi$ candidate
decay vertex.  To reduce backgrounds, all final state
particles are required to have $p_T > 0.5 \; \GeVc$, to be within
$\Delta R \equiv \sqrt{\Delta\eta^2 + \Delta\phi^2} < 1.0$ of each other,
and to originate within 5 cm of each other in $z$.
The signal modes are reconstructed identically to the reference modes,
except for the addition of two pions which originate from the $J/\psi$
candidate decay vertex.
The invariant mass of these two extra pions,
$m(\pipm)$, is required to be greater than $0.55 \; \GeVcc$.

The final vertex fit constrains the particles to originate from a
common vertex, except the $\Ks$ daughters whose combined momentum must
point back to that vertex.  For this fit, the invariant
masses of the $\Ks$ and $J/\psi$ candidates are constrained to
their world average measured values \cite{PDG}.
The $\chi^2$ of the fit is required to have a confidence level above $0.1\%$.

If there are multiple 
$B^0$ decay candidates in the same event, all candidates are kept.
Multiple candidates have the largest effect in the
$B^0 \to J/\psi \Kt \pipm$ sample where misassignments of the $K$ and
$\pi$ from the $\Kt$ result in a broad Gaussian shaped background.
Multiple candidates do not significantly affect the
$B^0 \to J/\psi \Ks \pipm$ sample.

To reduce background levels,
additional selection criteria are placed on the transverse
momentum of the neutral kaon, $p_T(K^{(*)0})$,
the proper decay time of the $B^0$ candidate, $ct(B)$,
and a $B^0$ isolation variable, $I \equiv p_T(B) / (p_T(B) + p_T(x))$.
The quantity $p_T(x)$ is the scalar sum of the transverse momenta of
all non-$B^0$ candidate tracks within $\Delta R < 1.0$ of the
$B^0$ candidate momentum direction.
For each signal mode these selection criteria are optimized to
maximize $S^2/(S+Bkg)$ of the signal sample
where $S$ is the expected signal size and $Bkg$ is the expected
background size.
This optimization uses the sidebands of the invariant mass distribution
of the signal data and the invariant mass distribution of the
reference data.

\begin{figure}[ht]
\begin{center}
\epsfig{file=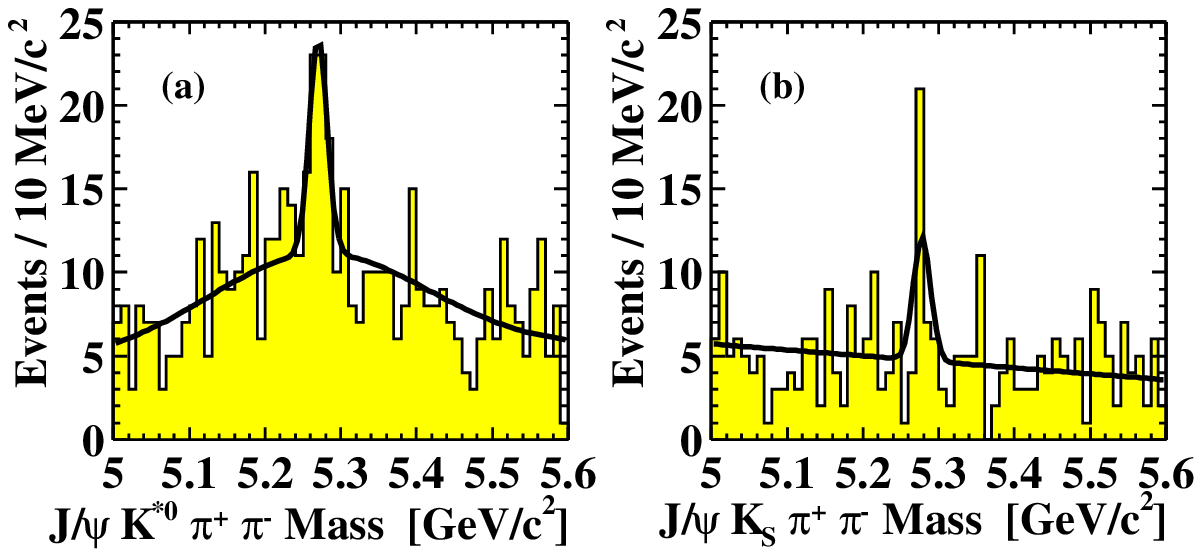, width=3.2in}
\end{center}
\caption{\label{fig:mass}
Invariant masses of $B^0$ candidates for $B^0 \to J/\psi \Kt \pipm$ (a)
and $B^0 \to J/\psi \Ks \pipm$ (b).}
\end{figure}

The optimized selection criteria for $B^0 \to J/\psi \Kt \pipm$ are
$p_T(\Kt) > 2.4 \; \GeVc$, $ct(B) > 170 \; \um$, and $I>0.60$.  
Figure \ref{fig:mass}(a) shows the resulting invariant mass peak.
The data are fit using the sum of a narrow and a broad Gaussian and
a linear background.  The width of the narrow Gaussian is fixed
to $11.8 \; \MeVcc$, based upon the expected width from the Monte Carlo
simulation scaled up by the ratio of widths between
the data and Monte Carlo simulation for the reference mode
$B^0 \to J/\psi \Kt$.
The fit results in $36.3 \pm 9.9$ signal events.
The $B^0 \to J/\psi \Kt$ reference mode with similar selection
criteria has
$257 \pm 18$ signal events. The ratio of reconstruction
efficiencies between $B^0 \to J/\psi \Kt$ and $B^0 \to J/\psi \Kt \pipm$ 
is $R_\epsilon = 3.75$.  These numbers lead to a ratio of branching
ratios of
$\BR(B^0 \to J/\psi \Kt \pipm) / \BR(B^0 \to J/\psi \Kt) = 0.53 \pm 0.15$,
where the error is statistical only.
Using $\BR(B^0 \to J/\psi \Kt) = 1.5 \times 10^{-3}$ \cite{PDG},
this corresponds to a branching ratio of
$\BR(J/\psi \Kt \pipm) = (8.0 \pm 2.2) \times 10^{-4}$.

The final state $J/\psi \Kt \pipm$ could come from the intermediate states
$B^0 \to \psi(2S) \Kt$ or $B^0 \to J/\psi \Kt \rho^0$.
Within $\pm2\sigma$ of the $B^0$ invariant mass shown in
Fig.~\ref{fig:mass}(a), there are 9 $\psi(2S) \to J/\psi \pipm$ candidates
on an expected background of 3
within $\pm10 \; \MeVcc$ of the $\psi(2S)$ mass of
3.686 $\GeVcc$ \cite{PDG}.
The remaining signal candidate events have higher $J/\psi \pipm$
invariant masses.
There is no identifiable resonant structure in the $\pipm$ invariant
mass distribution to indicate a large $\rho^0$
contribution, nor is there any identifiable resonant structure to the
$\Kt \pi^\pm$ or $\Kt \pipm$ invariant mass distributions.


The optimized selection criteria for the $B^0 \to J/\psi \Ks \pipm$ sample are
$p_T(\Ks) > 1.0 \; \GeVc$, $ct(B) > 30 \; \um$, and $I>0.50$.
This sample has less intrinsic background
than the $B^0 \to J/\psi \Kt \pipm$ sample
since the invariant mass peak of the $\Ks$ is narrower than that
of the $\Kt$ and its decay vertex is additionally
displaced from the $J/\psi$ decay vertex.
To take advantage of this lower background, events
that do not have enough SVX information to make
a precise $ct$ determination are included in a separate optimization
which does not restrict $ct(B)$.
The selection criteria for this sample are
$p_T(\Ks) > 1.9 \; \GeVc$ and $I > 0.70$.  These two samples are combined in
Fig.~\ref{fig:mass}(b).  A fit yields $21.0 \pm 6.3$ signal candidates.

The signal width is fixed in the fit to $\sigma = 11.3 \; \MeVcc$,
based upon the expected width from the Monte Carlo simulation
scaled up by the ratio of widths between the data and
Monte Carlo for the reference mode $B^0 \to J/\psi \Ks$.
Allowing this width to float results in a fitted width which
is approximately half the expected width.
The excess of events at the $B^0$ mass is robust across a wide
range of selection criteria and is broader when other selections are applied.
The normalized mass distribution
$(m_B - 5.28 \; \GeVcc) / \sigma_{m_B}$ has $\sigma = 0.67 \pm 0.21$
and $16.2 \pm 5.6$
events.  The unusual narrowness appears to be primarily an artifact
of this particular set of selection criteria which optimized the
expected $S^2/(S+Bkg)$.  Varying the
width by $\pm 20\%$ affects the fitted signal size by less than $2 \%$.

The ratio of efficiencies, $R_\epsilon = 4.98$, combined with 
$84.1 \pm 9.9$ $B^0 \to J/\psi \Ks$ reference events leads to 
a ratio of branching ratios of
$\BR(B^0 \to J/\psi \Ks \pipm) / \BR(B^0 \to J/\psi \Ks) =
1.24 \pm 0.40$, where the error is statistical only.  Using
$\BR(B^0 \to J/\psi \Ks) = (8.9 \pm 1.2) \times 10^{-4}$ \cite{PDG}
leads to
$\BR(B^0 \to J/\psi \Ks \pipm) = (1.1 \pm 0.4) \times 10^{-3}$.

\begin{figure}[ht]
\begin{center}
\epsfig{file=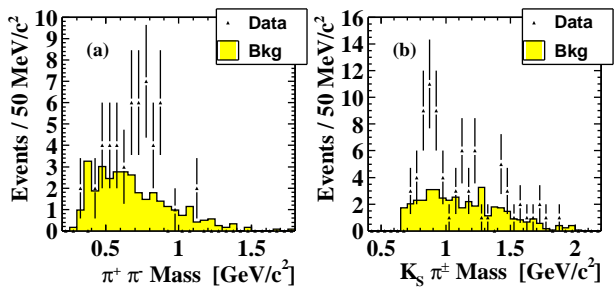, width=3.2in}
\end{center}
\caption{\label{fig:mpipi}
$m(\pipm)$ (a) and $m(\Ks \pi^\pm)$ (b)
for $J/\psi \Ks \pipm$ events within $\pm 2 \sigma$ of the $B^0$
invariant mass.}
\end{figure}

Unlike $B^0 \to J/\psi \Kt \pipm$, $B^0 \to J/\psi \Ks \pipm$ shows
evidence of several substructure contributions in addition to
$\psi(2S) \Ks$ candidates.
The $\pipm$ and $\Ks \pi^\pm$ invariant mass plots shown
in Fig.~\ref{fig:mpipi} have an excess of signal over background
in the $\rho^0$ and $K^{*\pm}$ invariant mass regions, indicating possible
contributions from $B^0 \to J/\psi \Ks \rho^0$ and
$B^0 \to J/\psi K^{*+} \pi^-$.
The backgrounds are estimated from the $\pipm$ and $\Ks \pi^\pm$ invariant
mass distributions of the candidates in the $B^0$ mass sidebands.

\begin{figure}[ht]
\begin{center}
\epsfig{file=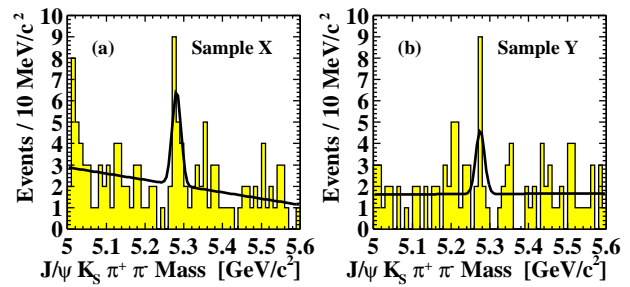, width=3.2in}
\end{center}
\caption{\label{fig:b_sub}
$B^0$ candidate invariant masses for events with a $K^{*\pm}$ candidate (a)
and those with a $\rho^0$ candidate but no $K^{*\pm}$ candidate (b).}
\end{figure}

To fit for these contributions, the $B^0$
invariant mass peaks for two samples of events are considered.
Events in sample $X$ have a $K^{*\pm} \to \Ks \pi^\pm$ candidate
with an invariant mass within $0.892 \pm 0.051 \; \GeVcc$.
Sample $Y$ contains events which have a $\rho^0 \to \pipm$
candidate with an invariant mass within $0.770 \pm 0.150 \; \GeVcc$
while excluding events in sample $X$.
The $m(\pipm) > 0.55 \; \GeVcc$ requirement is not placed upon
sample $X$.
Neither sample has any $\psi(2S)$ candidates.
The invariant mass peaks of the $B^0$ candidates in these samples
are fitted using a Gaussian signal of fixed width and a linear background.
Figure \ref{fig:b_sub} shows the results with sample $X$ having
$12.5 \pm 4.6$ fitted signal events and sample $Y$ having
$8.5 \pm 3.8$ fitted signal events.

Within $\pm 2 \sigma$ of the $B^0$ mass, sample
$X$ has 21 events on a background of 9.2; sample $Y$ has
14 events on a background of 7.3.
These numbers of events lead to Feldman-Cousins
$95\%$ confidence intervals \cite{FeldmanCousins} for the signal size
of [4.1, 22.6] and [1.1, 15.6] for samples $X$ and $Y$, respectively.
Using the fitted number of signal events and
the efficiencies for $B^0 \to J/\psi \Ks \rho^0$ and
$B^0 \to J/\psi K^{*+} \pi^-$ for each of the samples,
the resulting branching ratios are
$\BR(B^0 \to J/\psi \Ks \rho^0) = (5.8 \pm 3.1) \times 10^{-4}$ and
$\BR(B^0 \to J/\psi K^{*+} \pi^-) = (8.3 \pm 4.4) \times 10^{-4}$
with a correlation coefficient of -0.43.
These branching ratios assume that these two modes are the dominant
contributions to the two samples and that they do not interfere in
the overlap region of their $\Ks\pipm$ Dalitz plot.

\begin{figure}[ht]
\begin{center}
\epsfig{file=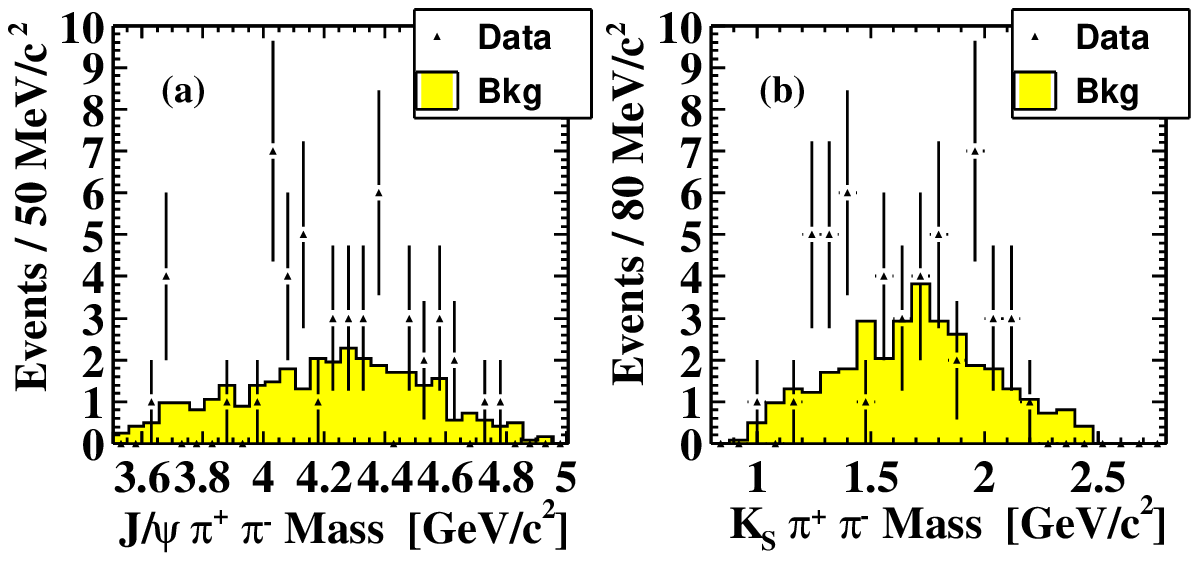, width=3.2in}
\end{center}
\caption{\label{fig:kspipi_sub}
$m(J/\psi \pipm)$ (a) and $m(\Ks \pipm)$ (b)
for $J/\psi \Ks \pipm$ events within $\pm 2 \sigma$ of the $B^0$
invariant mass.}
\end{figure}

The $J/\psi \pipm$ invariant mass plot of
Fig.~\ref{fig:kspipi_sub}(a) shows 4 $\psi(2S)$ candidates
on an expected background of 0.3.
It is possible that
$B^0 \to J/\psi \Ks \rho^0$ and $J/\psi K^{*+} \pi^-$ come from
$B^0 \to J/\psi K_1(1270)$.
Figure \ref{fig:kspipi_sub}(b) shows an excess in the $\Ks \pipm$
invariant mass distribution near $K_1(1270)$
but there is also a small excess of events at higher
$\Ks \pipm$ invariant masses.
The backgrounds are estimated from the $J/\psi \pipm$
and $\Ks \pipm$ invariant mass distributions of the candidates
in the $B^0$ invariant mass sidebands.

The results are summarized in Table \ref{tab:results}.
The dominant uncertainty in these branching ratios is the
statistical uncertainty due to the small signal size.  Many of
the systematic uncertainties cancel in the ratio of branching ratios
with the reference mode.  The systematic uncertainties that
do not cleanly cancel are summarized in Table \ref{tab:uncertainties}
and described as follows.

\setcounter{topnumber}{10}
\setcounter{bottomnumber}{10}
\setcounter{totalnumber}{10}
\setcounter{dbltopnumber}{10}

\renewcommand{\topfraction}{0.9}
\renewcommand{\dbltopfraction}{0.9}
\renewcommand{\bottomfraction}{0.9}
\renewcommand{\textfraction}{0.1}
\renewcommand{\floatpagefraction}{0.9}
\renewcommand{\dblfloatpagefraction}{0.9}

\begin{table*}[h!tb]
\begin{tabular}{l|llll}
\hline
\hline
	& $J/\psi \Kt \pipm$ & $J/\psi K^0 \pipm$
	& $J/\psi K^0 \rho^0$ & $J/\psi K^{*+} \pi^-$ \\
\hline
$N_{obs}$	& 85	& 39		& 14		& 21	\\
Bkg		& 54.0	& 21.1		& 7.32		& 9.22	\\
$S_{fit}$	& $36.3 \pm 9.9$	& $21.0 \pm 6.3$
		& $8.5 \pm 3.8$		& $12.5 \pm 4.6$	\\
$S_{FC}$	& [13.9, 50.8]	& [7.1, 31.8]	& [1.1, 15.6]	& [4.2, 22.6]	\\
$S_{ref}$	& $257 \pm 18$		& $84.1 \pm 9.9$
		& $84.1 \pm 9.9$	& $84.1 \pm 9.9$	\\
$R_\epsilon$	& 3.75	& 4.98		& ---		& ---	\\
BR		& $(8.0 \pm 2.2 \pm 1.5) \times 10^{-4}$
		& $(1.1 \pm 0.4 \pm 0.2) \times 10^{-3}$
		& $(5.8 \pm 3.1 \pm 1.2) \times 10^{-4}$
		& $(8.3 \pm 4.4 \pm 1.7) \times 10^{-4}$	\\
\hline
\hline
\end{tabular}
\caption{\label{tab:results}
Summary of results: Number of observed events ($N_{obs}$);
Fitted background (Bkg);
Fitted signal ($S_{fit}$);
Feldman-Cousins $95\%$ confidence interval on the signal size ($S_{FC}$);
Fitted number of reference mode signal events
($S_{ref}$);
Ratio of efficiencies ($R_\epsilon$);
and the branching ratio (BR) where the first uncertainty is statistical and
the second is systematic.}
\end{table*}

\begin{table}[ht]
\begin{center}
\begin{tabular}{l|ll}
\hline
\hline
$\BR(B^0 \to J/\psi K \pipm)$ 	& $\%$ Uncertainty & \\
Source of Uncertainty	& $K = \Kt$	& $K = K^0$	\\
\hline
Reference Mode BR	& 11	& 14	\\
Helicity Model		& 9.9	& 9.4	\\
Signal Width		& 7.5	& ---	\\
Trigger Model		& 5.0	& 5.0	\\
Monte Carlo Composition	& 5.0	& 5.0	\\
$B^0$ Production Model	& 2.5	& 2.5 	\\
\hline
Combined Uncertainties	& 18	& 18	\\
Without Ref. Mode BR	& 15	& 12	\\
\hline
\hline
\end{tabular}
\end{center}
\caption{\label{tab:uncertainties}
Systematic uncertainties on $\BR(B^0 \to J/\psi K^{(*)0} \pipm)$.}
\end{table}

The uncertainty on the reference mode branching ratio does not enter
into the ratio of branching ratios but it is the dominant uncertainty
for the branching ratio measurements.  It is $11\%$ for $J/\psi \Kt$ and
$14\%$ for $J/\psi \Ks$.
The signal modes have more helicity degrees of freedom than the
reference modes and the relative contributions of possible helicity
states are not known.
This introduces an uncertainty in the efficiencies modeled with
the Monte Carlo simulation
of $9.9\%$ for $J/\psi \Kt \pipm$ and $9.4\%$ for $J/\psi \Ks \pipm$.
There is a $5\%$ uncertainty in both signal modes due to uncertainties
in the trigger model used in the Monte Carlo.
Multiple decay modes could contribute to the final states studied here but
they all have similar reconstruction efficiencies.  Varying the relative
compositions in the Monte Carlo results in a net uncertainty of $5\%$.
The effect of differing $p_T(B)$ spectra from various $B$
production models introduces a $3\%$ uncertainty.
Varying the fitted signal width and the width of the
broad Gaussian shaped background by $\pm 10\%$
in the $B^0 \to J/\psi \Kt \pipm$ sample results in a $7.5\%$
variation in the fitted signal size and is thus included as
a systematic uncertainty.  Varying the signal width by $\pm 20\%$
in the $B^0 \to J/\psi \Ks \pipm$ sample has less than a $2\%$
effect and thus is neglected.
The $J/\psi \Ks \rho^0$ and $J/\psi K^{*+} \pi^-$
branching ratios assume no interference in their overlap region in
the $\Ks \pipm$ Dalitz plot.  Completely constructive interference
would increase their combined branching ratio by $\sim 20\%$;
a $10\%$ systematic uncertainty is included in each mode to account
for this possibility.


The most statistically significant mode, $J/\psi \Kt \pipm$,
has a significance of $3.7 \sigma$.
$J/\psi \Ks \pipm$ has a significance of $3.3\sigma$;
its submodes $J/\psi \Ks \rho^0$ and $J/\psi K^{*+} \pi^-$
show hints of a signal but have less than $2\sigma$ significance.
The measured branching ratios are
large enough that these modes should be visible in the data
already recorded by the CLEO, Belle, and BaBar experiments.
CDF should record hundreds of these decays in Run II.

We thank the Fermilab staff and the technical staffs of the participating
institutions for their vital contributions.  This research was supported
by the U.S. Department of Energy and the National Science Foundation;
the Italian Istituto Nazionale di Fisica Nucleare; the Ministry of Education,
Science, Sports and Culture of Japan; the Natural Sciences and Engineering
Research Council of Canada;
the National Science Council of the Republic of China;
the Swiss National Science Foundation; the A.~P. Sloan Foundation;
the Bundesministerium f\"ur Bildung und Forschung, Germany;
the Korea Science and Engineering Foundation (KoSEF);
the Korea Research Foundation;
and the Comision Interministerial de Ciencia y Tecnologia, Spain.




\begin{thebibliography}{9}
\bibitem{CDF_coords}
At CDF, the cylindrical axis of the CTC defines the $z$ axis, which is aligned
along the proton beam direction.  $\phi$ is the azimuthal angle,
$\theta$ is the polar angle, and $r$ is the radial distance from the $z$
axis.  The pseudorapidity is defined as $\eta = -\ln(\tan(\theta/2))$.
\end{thebibliography}
\end{document}